\documentclass[11pt,a4paper]{article}
\pdfoutput=1
\usepackage{jheppub}
\usepackage[T1]{fontenc}
\usepackage{ulem}
\usepackage{float}
\usepackage{cancel}

\def\be{\begin{equation}}
\def\ee{\end{equation}}
\def\beq{\begin{eqnarray}}
\def\eeq{\end{eqnarray}}
\def\bn{\begin{eqnarray*}}
\def\en{\end{eqnarray*}}

\def\p{\phi}
\def\w{\omega}

\def\b{\beta}

\def\g{\gamma}

\def\G{\Gamma}

\def\pd{\partial}
\def\e{\epsilon}

\def\m{\mu}
\def\r{\rho}

\def\ket{{\rangle}}

\def\l{\lambda}

\def\cA{{\cal{A}}}

\title{Generalized Landau Yang Theorem}

\author[a,b]{T. R. Govindarajan,}
\author[c]{Rakesh Tibrewala}

\affiliation[a]{The Institute of Mathematical Sciences,\\Chennai 600113, Tamil Nadu, India}
\affiliation[b]{Krea University,\\Sricity, 517646, Andhra Pradesh, India}
\affiliation[c]{Department of Physics, The LNMIIT,\\ Jaipur 302031, Rajasthan, India}

\emailAdd{trg@imsc.res.in, govindarajan.thupil@krea.edu.in}
\emailAdd{rtibs@lnmiit.ac.in}

\abstract{Landau Yang theorem is well known for the past several decades.
It prohibits the decay of a massive spin 1 particle to two photons.
This emerges simply from the representation theory of the Poincare group
and Bose Statistics. 
It  does not require any action or Lagrangian. We generalize this theorem to 
theories with supersymmetry (SUSY) which disallows even 
decay to two photinos (Majorana fermions) as well as the decay of a zino to a photon and a photino. We will prove that
if the photon has a mass, howsoever small, this theorem 
can be evaded. We also show that the supersymmetric selection rule above can also be evaded
through the Stueckelberg mass term. Further interesting implications are 
also pointed out.}

\begin{document} 
\maketitle
\flushbottom

\section{Introduction}
Landau \cite{landau}  and Yang \cite{yang} independently proved a 
theorem which disallows a massive 
spin 1 particle from decaying to two massless spin-1 bosons
without further internal symmetries. 
The elegance of the theorem stems from the fact that it does 
not depend on any interactions and is very generic. This is 
easy to understand from the representation theory of the Poincare 
group and Bose-Einstein statistics. For example, the
$\r$ meson which has spin 1 does not decay to two photons whereas the pion (which has spin 0) does.
It prohibits the eletroweak $Z$ particle to decay to two photons (the bound on the branching ratio $\G_{Z\rightarrow\g\g}/\G_{\mathrm{Tot}} < 5.2\times 10^{-5}$ \cite{PDG-z2g}). 
Recently this theorem has been applied to understand that the
Higgs Boson cannot be a spin 1 particle due to its decay to two photons \cite{Atlas, CMS}. 
 
\subsection{Simple proof}
Massive spin 1 particle can be taken to be at rest with 
the four momentum $P^\m~=~(m,0,0,0)$. Its spin 
can be given as $\vec{S}$. The massless photons 
will have momenta $\vec{k}$ and $-\vec{k}$ in order to 
conserve the momentum. The polarization
of these two photons, denoted $\vec{s}_1,\vec{s}_2$, will be in the plane perpendicular 
to $\vec{k}$ so that  
\be \vec{k}\cdot \vec{s}_1 ~=~0~=~\vec{k}\cdot \vec{s}_2
\label{polarization}
\ee
Now let us write down all possible terms linear in $\vec{S},\vec{s}_1,\vec{s}_2$ that can appear in the amplitude for the decay of the massive spin 1 particle to two photons.
There are only three generic terms that can be written as 
candidates for the amplitudes. These are:
\begin{enumerate}
\item $A_1~=~a_1(k^2)(\vec{S}\cdot \vec{k})~(\vec{s}_1\cdot \vec{s}_2)$
\item $A_2~=~a_2(k^2)(\vec{s}_1\times \vec{s}_2)\cdot\vec{S}$
\item $A_3~=~a_3(k^2)(\vec{s}_1\times \vec{s}_2~\cdot \vec{k})~
(\vec{S}\cdot\vec{k})$
\end{enumerate}
Here $a_i(k^2)$ are arbitrary invariant functions.

The general form of the amplitude is 
\be
\cA(\vec{k},\vec{s}_1,\vec{s}_2,\vec{S})~=~\sum_1^3 A_i
\label{amp}
\ee
Now the Bose symmetry of photons implies that the amplitude 
should be the same if we transform $\vec{k} \rightarrow -\vec{k}$ 
and interchange $\vec{s}_1~\leftrightarrow~ \vec{s}_2$.
That is we should have
\be 
\cA(\vec{k},\vec{s}_1,\vec{s}_2,\vec{S})~=~
\cA(-\vec{k},\vec{s}_2,\vec{s}_1,\vec{S})
\ee
But $\cA(-\vec{k},\vec{s}_2,\vec{s}_1,\vec{S})~=~-
\cA(\vec{k},\vec{s}_1,\vec{s}_2,\vec{S})$ implying 
\be 
\cA(\vec{k},\vec{s}_1,\vec{s}_2,\vec{S})~=~0
\ee
The vanishing of the amplitude is generic 
and does not depend on the details of the action 
or the nature of discrete transformation property of the 
initial vector particle. We can see that the above theorem 
stems only from the representation theory of the Poincare 
group and Bose symmetry. 

\subsection{Representation theory of Poincare symmetry}
We recollect \cite{wigner} that massless 
irreducible representations of the Poincare 
group are  given by the helicity. On the other hand, massive 
representations are specified by the mass parameter as well as 
the spin value coming from the Pauli-Lubanski vector. We also need to 
represent the massless photons covariantly using a four vector $A_\m$
to discuss further interactions through gauge theory. 
That brings in the helicity $\pm ~1$. These follow from the 
induced representations of the 
little group of four momentum vector 
$p^\m~=~(\w,0,0,\w)$ of a massless particle which 
is the Euclidean group in two dimensions.   
The little group for a massive particle on the other hand 
is $SO(3)$ which comes  from
the fact that the massive particle can be brought to rest in a frame
and its four momentum in that frame is $(m,0,0,0)$. 
In addition it is well known that the helcity $\pm~1$ representations
from $A_\m$ are reducible but not decomposable \cite{mathews}.
We need to look at the content of the product of two such helicity
representations through Clebsch Gordon coefficients. Also   
Bose symmetry should be imposed. It is clear that for a massive spin 1 particle to decay to two photons demands that the symmetric 
product of two massless representations of the Poincare 
group contains a massive 
spin one representation. It was shown by 
Balachandran and Jo \cite{bal} 
that it does not contain such a representation.    
The direct product of two massless representations with 
four momenta $(p,0,0,\pm p)$ and helicities $\l_1,\l_2$
can be given by:
\be 
|\l_1~\l_2~\hat{p}~j~\mu\ket = ~\int_{SU(2)}~d\xi(R) D_{\mu~\l_1-\l_2}^j|p\l_1\ket~|-p\l_2\ket
\ee
where $d\xi(R)$ is the Haar measure on $SU(2)$ and $D_{\mu~\l_1-\l_2}^j$ are the spin-j
representation matrices. For the case of two identical  
particles we have the (anti)symmetric combinations
\be
|\l_1~\l_2~\hat{p}~j~\m\ket_{s,a}~=~\frac{1}{2}\left(|\l_1~\l_2~\hat{p}~j~\m\ket~\pm~
(-1)^{j+\l_1+\l_2}~|\l_2~\l_1~\hat{p}~j~\m\ket~\right)
\ee
where the subscripts $s, a$ correspond to the upper and lower signs on the rhs respectively. For photons obeying Bose statistics we need the symmetric product.
For the symmetric product with two identical particles of helicities $\l_{i}=\pm 1, i=1,2$ we have for $\l_1=\pm\l_2$: 
\be
|\l_{1}~\l_{2}~\hat{p}~j~\m\ket~_s~=~\frac{1}{2}(1~+~(-1)^{j+\l_{1}+\l_{2}})~|\l_{1}~\l_{2}~\hat{p}~j~\m\ket~
\label{product}
\ee
With $j=1$ we find the right hand side vanishes. This implies 
two photon states cannot have $j=1$ component thereby forbidding, for example, 
$Z \rightarrow 2\g$ transition and many others (in \cite{choi} the helicity formalism is used to generalize the theorem for the decay to two identical massless particles of any spin). On the other hand 
if $j=0$ as we expect for the Higgs,
the decay to two photons is not prohibited by the theorem and provides 
a strong evidence for the scalar nature of Higgs \cite{Atlas,CMS}.

\section{Generalization  to SUSY}
We will now study the extension of the above theorem in the
Supersymmetric context. The obvious starting point 
is the representation theory of the Super Poincare algebra. 

Super Poincare algebra also has two Casimir operators. The first 
is the usual $P_\m~P^\m$ corresponding to the mass of the particles. 
But the spin from the Pauli Lubanski vector does not 
provide  the other invariant. There is a generalization of this, known as the
superspin operator which commutes with all the generators of the super Poincare algebra \cite{buchbinder}. 
The mass $m$ and the superspin operator $Y$ provide the 
massive irreducible representations. When Y = 0 we get a supersymmetric multiplet consisting of
a scalar and a spin $1/2$ particle of the same mass.  

The vector multiplet $V$ (obtained for $Y=1/2$) 
consists of a massive vector particle
(Z), two chiral spinors (zinos) and a scalar ($\p$). On the 
other hand, for the massless representation of photon $\g$ we have only
the Majorana particle photino $\tilde{\g}$ as the fermionic partner.

Given the above particle content we can easily see the extension
of the Landau Yang theorem to forbid decay of $Z \rightarrow \g + \g$.
Not only that, in the effective field theory we will not have any interaction
term representing the three superfields whose content includes
$Z,\g,\g$ as external legs.

\begin{figure}[H]
\centering
\includegraphics[scale=0.5]{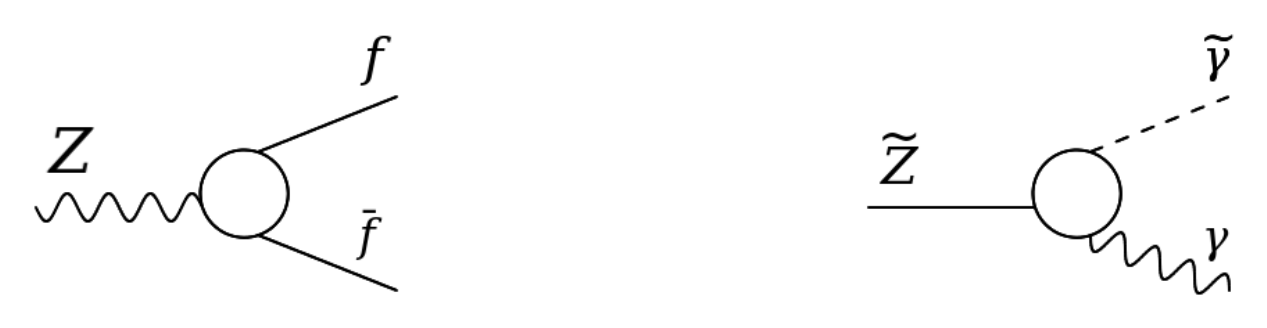}
\caption{Diagrams showing (a) the decay of $Z$ to two massless Majorana fermions (left) and (b) the decay of the zino to a photino and a photon (right)}
\label{zdecays}
\end{figure}

Hence combined with supersymmetry we will also have in addition
zero amplitude for Z decaying to two photinos. Here the Fermi
symmetry of photinos along with SUSY forbids such a process.
Interestingly such a decay is prohibited by simply Poincare group
representation theory itself without requiring SUSY -- a massive spin 1 particle cannot
decay to two (identical) massless Majorana particles, figure \ref{zdecays}(a). This was shown in \cite{Bal-Book} and is seen
in Balachandran and Jo paper through the decomposition of
the antisymmetric product  of two spin $1/2$
Majorana representations. To see this consider eq. (\ref{product})
with $j=1, \l_{i} = 1/2$ corresponding to decay to two photinos. 
It is easy to see that the right hand side of eq.\ref{product} is 
identically zero.

A more interesting conclusion and which explicitly requires supersymmetry is
the zero amplitude for
a zino going to a photon and a photino, figure \ref{zdecays}(b). This is inspite of the fact that all
three particles are distinct. Only representation theory
of the super Poincare algebra along with symmetric/antisymmetric product of irreducible
reprsentations provides such a powerful result. 

\section{Evading the theorem}
In the literature several options have been discussed for 
evading  the theorem through relaxing the conditions for the theorem to hold.
For example, one can consider  a background of 
homogeneous electric or magnetic fields \cite{rindani} 
there by relaxing the requirement of Lorentz 
invariance/Poincare symmetry. The external field gives another 
vector in addition to the polarization vectors and which therefore allows 
nonzero amplitude for the process. Such background magnetic 
fields are present in astrophysical contexts. If the 
massless particles are gluons then there are additional internal 
color degree of freedom which gives scope for decay of 
$Z \rightarrow g~g$ in QCD \cite{qcd, beenakker} (but the caveat is that gluons are always confined). 

Much more recently, using noncommutative  
geometry which is considered as violating 
Lorentz symmetry, the 
possibility  of $Z \rightarrow \g + \g$ has been considered \cite{behr}.
It is well known that massless nature of photon leads to infrared 
divergences in QED. This was studied extensively in the sixties 
and was related to the difficulties associated with gauge 
invariant description of electron which is always surrounded 
by soft photons (photons of infinitesimal frequencies). This led 
to a new possibility to evade the Landau-Yang theorem 
arising from the fact that due to the asymptotic 
properties of QED the contribution of soft photons can be 
included in a way which provides Lorentz noninvariant sectors \cite{balnew}.

\section{Massive photons and Landau-Yang theorem} 

Unlike some of the above cases, we consider an alternative which does not require breaking of Lorentz invariance. We will consider the photons to be massive (note that experimentally there is only an upper bound on the mass of the photon, see subsection \ref{Inonu-Wigner} below). 
It is obvious that if the photons have a small mass the question 
of the representation theory of the Poincare group takes a new angle.
It becomes a question of whether  
in the tensor product of two massive spin 1 photons obeying 
Bose symmetry there is a massive spin 1 representation 
(Z). The answer is obviously yes. This possibility 
arises because massive photons have three degrees of freedom
including the longitudinal photons and eq.\ref{polarization} 
is no longer valid. It gives a longitudinal 
component to the polarization $\vec{s}_i$ in the direction of $\vec{k}$ so that $k\cdot\vec{s}_i \neq 0$. 
This allows for a nonzero amplitude for the decay of a massive spin 1 particle
like $Z$ to two massive $\g$ with a subtantial contribution from the 
longitudinal component for one of the photons.

\subsection{Inonu-Wigner contraction} \label{Inonu-Wigner}
We mentioned that if the photon has a mass, even if it is small,
the Landau-Yang theorem can be evaded. For this one needs to understand the 
nature of the limit $m_\g~\rightarrow 0$.  
The question of mass of the photon was raised by Schrodinger 
himself in 1955 \cite{schrodinger} and answered by him  
that QED results will not be 
affected by a small photon mass as long as its interaction with
matter is only through a conserved current, that is, if
$L_{int}~=~j_\m~A^\m$ where $\pd_\m~j^\m~=~0$. 
Schrodinger himself worked out the upper bound of the mass of the photon 
using geomagnetic data.  
The current limit for the mass of the photon 
as given by the Particle data book \cite{PDG} is $m_\g~<~10^{-18}eV$. 
The question of mass going to zero limit can be posed as group 
theoretic question of contraction of $SO(3)~\rightarrow E(2)$. The contraction of the massive representation of the Poincare group to the massless representation is discussed in Kim and Wigner \cite{kim} and here we briefly summarize the parts relevant for us.
The generators of $SO(3)$ can be written as $L_\pm,L_3$
with the Lie-algebra
\be
\left[L_3,L_\pm\right]~=~\pm L_\pm,\qquad [L_{+},L_{-}]~=~2L_3
\ee
On the other hand the group $E(2)$ is specified by $P_x,P_y,L_3$
with the algebra
\be
[P_x,P_y]~=~0,\qquad [L_3,P_{x,y}]~=~\pm i~P_{y,x}
\ee
The contraction $SO(3)~\rightarrow E(2)$ in the limit $m\rightarrow 0$ was provided by Inonu and  Wigner 
through the mapping 
\be P_{x,y}~=~\pm \frac{1}{R}S^{-1}L_{x,y}~S~\qquad S~=~\begin{pmatrix}1&0&0\\
0&1&0\\ 0&0&R \end{pmatrix}
\ee
In the $\e~=~\frac{1}{R} \longrightarrow 0$ we get $E(2)$.
For all values $\e \neq 0$ it provides the same group $SO(3)$.
When $\e~=~0$ we get the sphere of infinte radius 
close to the northpole and rotations about $x$ and $y$ axes are 
related to translations along $y$ and $x$ axes, respectively.

Both $SO(3)$ and $E(2)$ are 
subgroups of the Lorentz
group and we should arrive at this contraction as part of $SL(2,C)$ \cite{kim}.
This is explained better by looking at the rotation and
boost generators $J_i,~K_i$ and considering 
\be N_1~=~K_1~-~J_2,~~N_2~=~K_2~+~J_1.
\ee
Then $J_3,N_1,N_2$ give the Euclidean group $E(2)$.
For this we define the light cone coordinates $z_{\pm}~=~
z~\pm~t$. In the light cone coordinates we define:
\be 
S(R)~=~\begin{pmatrix}1&0&0&0\\
0&1&0&0\\ 0&0&R&0 \\ 0&0&0&\frac{1}{R}\end{pmatrix}
\ee
where $R~=~\sqrt{(1+\b)\over (1-\b)}$ and $\b$ is the velocity and the generators
$N_i~=~-\frac{1}{R}\e_{ij}~S~J_j~S^{-1}$ where the $J_{i}$ are the angular momentum operators in the light cone coordinates.

At the level of the Poincare group this amounts to 
studying massless representations as the limit of 
massive representation. Specifically, for the spin 1 particle,
the extra degree of freedom that the vector potential  
gets in the massive case (known as the Stueckelberg degree of freedom) interacts 
with matter only as a surface term in the massless limit. 

\section{The process: Stueckelberg Weinberg-Salam model}
\begin{figure}
\centering
\includegraphics[scale=.7]{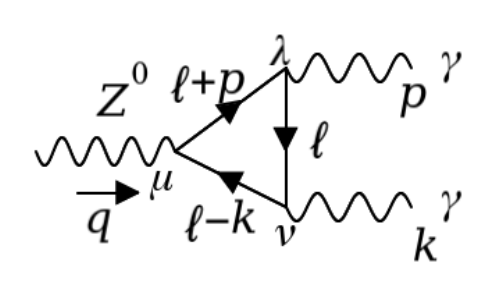}
\caption{$Z \rightarrow 2 \g$ loop diagram }
\label{z2g}
\end{figure}

It is well known that simply adding a mass term to the gauge boson Lagrangian violates the gauge invariance of the theory. To circumvent this we will consider the Stueckelberg extension of the abelian gauge boson in the Weinberg-Salam model \cite{stueckelberg,ruegg}.
As an explicit example we consider the $Z\rightarrow 2\g$ process. $Z$ boson being chargeless is not 
coupled  directly to the photon. We have to look for loop diagrams. 
Z interacts with the electron and the electron is 
coupled to the photon. The one loop diagram figure \ref{z2g}
is an example of two photon emission from $Z$ (there will be another diagram with $(p,\nu)\leftrightarrow (k,\lambda)$ which we do not show explicitly). 
In principle we will have contributions from the muon ($\mu$) and tau ($\tau$) loops as well as quark loops. Their 
contributions are expected to be small. 
The new amplitude will have an extra term $A_4$ in addition to those in
eq.\ref{amp}, where $A_4$ is given by:
\be
A_4~=~(\vec{s}_{1}\cdot\vec{S})(\vec{s}_{2}\cdot\vec{k})-(\vec{s}_{1}\cdot\vec{k})(\vec{s}_{2}\cdot\vec{S})
\ee
This, as mentioned above, is because we no longer have the transversality 
condition  $\vec{k}\cdot \vec{s}_i~=~0$. 

Ignoring the coupling and the external gauge bosons, contribution from diagram with only electron in the loop is 
\be
T \approx (-1)\int\frac{d^{4}\ell}{(2\pi)^{4}}Tr\left[\gamma^{\mu}\frac{(1-\gamma^{5})}{2}\frac{i}{C_{1}}\gamma^{\lambda}\frac{i}{C_{2}}\gamma^{\nu}\frac{i}{C_{3}}\right] + (p,\nu)\leftrightarrow (k,\lambda)
\ee
where
\bn
C_1~&=&~\cancel{\ell}~-~\cancel{k}~-~m_e \\
C_2~&=&~\cancel{\ell}~-~m_e \\
C_3~&=&~\cancel{\ell}~+~\cancel{p}~-~m_e
\en
This has to be contracted with the massive gauge boson polarization vectors. As has been pointed out in \cite{ruegg}, the Stueckelberg extension of the standard model is anomaly free and therefore we can simply estimate the amplitude for the process without worrying about the details of evaluating the loop integral. Including the coupling constants and the masses of the particles involved, we can estimate $|T|^{2}$ (upto numerical factors) to go as
\be
|T|^{2} \approx (v^{Z}_{e})^{2}(v^{A}_{e})^{4}M_{Z}^{2}\left(\frac{m_{\g}}{m_{e}}\right)^{2}.
\ee 
In the above expression $v^{Z}_{e}$ and $v^{A}_{e}$ are, respectively, the effective couplings of the electron with the $Z$ boson and the photon (in the notation of \cite{ruegg}), $M_{Z}$ is the mass of the $Z$ boson, $m_{e}$ is the mass of the electron and $m_{\g}$ is the mass of the photon. We would like to point out that there are differences between the conventional Weinberg Salam model and the Stueckelberg extension\footnote{Here we have projected the $Z-e-e$ vertex to the $V-A$ part since we are only interested in the order of magnitude estimate for the process. However, we note that in the Stueckelberg extension of the standard model there is a slight difference in the electromagnetic coupling of the photon to the left-handed and the right-handed electrons and we refer the reader to the review \cite{ruegg} for further details.}. We see that in the limit $m_{\g}\rightarrow 0$ the $Z\rightarrow\g\g$ does not occur as required by the Landau-Yang theorem.


Even though for massive photons we have a non-zero contribution, observing such an effect experimentally is difficult because
\begin{enumerate}
\item since $m_{\g}$ is very small, the probability for the decay $|T|^{2} \propto M_Z^{2}(m_{\g}/m_e)^{2}$ is itself very small. Observing such contributions at present is extremely difficult.
\item Observing longitudinal photon by itself is very difficult 
as its interaction with matter is very small. The only option is 
the missing energy and momentum carried by it.
\end{enumerate}
In spite of the observational difficulties, a non-zero photon mass, howsoever small, allows for escaping the Landau-Yang theorem 
without breaking Lorentz/Poincare symmetry which has been the 
contention of all the other mechanisms.
We will comment on issues in connection with the violation of the 
Landau-Yang theorem in conclusions. 

\subsection{Stueckelberg extension of SUSY}  
Lastly, we can extend the above ideas and consider the Stueckelberg extension of supersymmetric QED interacting with a massive vector supermultiplet by providing mass to the photon and its superpartner the photino. For this we need a chiral multiplet in addition to the vector multiplet (for the photon) to cancel the 
supersymmetric gauge transformations. The chiral multiplet is the supersymmetric analogue of the Stueckelberg field. 
We refer to the literature for such an extension \cite{buchbinder,nath,trg}. 

\section{Conclusions}
Landau Yang theorem is an important selection rule which has 
helped in understanding the role of symmetries in Particle Physics. While the original work \cite{landau, yang} showed the impossibility of the decay of a massive spin one particle to two identical massless bose particles, the conclusion is true even for identical massless fermions of the same helicity \cite{bal, Bal-Book}. In the present work we have also provided extension of the theorem to supersymmetry which, very interestingly, prohibits the decay of a massive spin one-half particle to even two non-identical massless particles which, nevertheless, are part of the same supersymmetric multiplet. In particular, we showed that in the supersymmetric extension of the SM, the zino cannot decay to a photon and a photino.

As the Landau Yang theorem is related to massless particles it has acquired 
importance coming from asymptotic properties of fields. 
It has also made the understanding of the Hilbert space of states 
with appropriate properties the central focus. In this connection 
the mass of the particle as a Lorentz covariant regulator for infrared questions 
is crucial. Similarly semi-inclusive cross sections for S-matrix in scalar QED, for instance, seem to require mass for the photon as a regulator to remove ambiguities in handling infrared divergences \cite{laddha}. As seen, the theorem can be evaded when the photon has a nonzero mass. This is achieved without violating the Lorentz invariance by using the Stueckelberg mechanism as opposed to some of the previous studies where the theorem could be evaded only at the cost of Lorentz invariance \cite{rindani, qcd, balnew}. 

It is easy to understand the role of mass as a regulator for this theorem.
The process $Z\rightarrow 2\gamma$ can come from a gauge invariant term of the form
$S_{int}~=\int d^{4}x ~g(F_{\mu\rho}F_{\nu\sigma}g^{\rho\sigma})(\partial^\mu~V^\nu)$ where $V^{\mu}$ is the massive spin one vector field and $F_{\mu\nu}$ is the electromagnetic field strength. Integrating by parts leads to $\partial_\mu~F^{\mu\nu}$ which is zero onshell. If the photon has a small mass, it will provide a nonzero result. The same is true even for supersymmetric theories as well where there will be an additional term in $S_{int}$ given by $\psi^{T}[\gamma_{\mu},\gamma_{\nu}]\psi\partial^{\mu}V^{\nu}$ corresponding to the photino. Once again, in the massless theory this contribution will be zero on-shell as can be confirmed by integrating by parts implying that $Z\rightarrow\tilde{\gamma}\tilde{\gamma}$ is also not allowed. This is consistent with what has been said earlier regarding the decay of a massive spin one particle to two Majorana fermions. But with broken supersymmetry, the photino will become massive and the decay of $Z$ to two photinos becomes allowed. Phenomenologically this is interesting since such decays, if observed, can provide important information about the scale of SUSY.

As Schrodinger \cite{schrodinger}  pointed out it is important 
to understand the approach to masslessness of particles 
even though nature may require zero mass. 
\vskip.5cm
\emph{Even if we find in Nature the limiting case is realized, we should still feel the urge to adumbrate a theory which agrees}
\emph{with experience on approaching to the limit, not by a sudden.} \hfill{E Schrodinger}
\vskip.5cm

A small photon mass may provide 
some novel ways of understanding some cosmological questions on the nature of dark matter/energy \cite{trgdark} as well.  
Higgs mechanism through which photon can acquire mass may also provide a link between
primordial blackholes and dark matter in 
early galaxies \cite{isra,SMBH}. It is expected that at length scales $L~\leq~\frac{h}{m_\g}$
the longitudinal photon behaves like a Stueckelberg scalar. But experimental observation 
of emission of such a scalar is very difficult if the mass is small as explained 
by Schrodinger \cite{schrodinger}. The only observational signature would be missing energy and momentum in the process
$Z\longrightarrow \g_T~+~\g_L$ (which could be a fraction of the energy-momentum carried by the transverse photon, although the rate of such processes will be limited by $M_{Z}(m_{\g}/m_{e})^{2}$, as seen earlier).

\acknowledgments

TRG would like to thank Alok Laddha and K.~S.~Narain for discussions. Authors would like to thank Parmeshwaran Nair for his comments on the draft. Feynman diagrams were made using Feynman Diagram Maker (https://www.aidansean.com/feynman/).

\end{document}